\documentclass[10pt,onecolumn]{elsarticle} 

\usepackage{graphics,graphicx}
\usepackage{amsmath,amsfonts,amssymb,mathtools}
\usepackage[dvipsnames]{xcolor}
\usepackage{hyperref}

\newcommand{\xb}{\mathbf{x}}

\newcommand{\zb}{\mathbf{z}}
\newcommand{\fb}{\mathbf{f}}
\newcommand{\hb}{\mathbf{h}}
\newcommand{\vb}{\mathbf{v}}

\newcommand{\wb}{\mathbf{w}}
\newcommand{\Ab}{\mathbf{A}}
\newcommand{\Xb}{\mathbf{X}}
\newcommand{\Cb}{\mathbf{C}}
\newcommand{\Wb}{\mathbf{W}}
\newcommand{\Zb}{\mathbf{Z}}
\newcommand{\Qb}{\mathbf{Q}}
\newcommand{\Ib}{\mathbf{I}}
\newcommand{\Fb}{\mathbf{F}}

\newcommand{\Pb}{\mathbf{P}}
\newcommand{\muB}{\mbox{\boldmath$\mu$}}

\def\one{\scriptscriptstyle{1}}
\def\zero{\scriptscriptstyle{0}}

\begin{document}

\begin{frontmatter}

\title{Target tracking in the framework of possibility theory: The possibilistic Bernoulli filter}

\author[RMIT]{Branko Ristic\corref{cor1}}
\ead{branko.ristic@rmit.edu.au}
\cortext[cor1]{Corresponding author: B. Ristic, RMIT University, GPO Box 2476, Melbourne VIC 3001, Austalia; tel: +61 3 9925 3302.}

\author[War]{Jeremie Houssineau}
\ead{jeremie.houssineau@warwick.ac.uk}

\author[DSTG]{Sanjeev Arulampalam}
\ead{sanjeev.arulampalam@dst.defence.gov.au}

\address[RMIT]{RMIT University, Australia}
\address[War]{University of Warwick, United Kingdom}
\address[DSTG]{Defence Science and Technology Group, Australia}

\begin{abstract}
The Bernoulli filter is a Bayes filter for joint detection and tracking of a target in the presence of false and miss detections. This paper presents a mathematical formulation of the Bernoulli filter in the framework of possibility theory, where  uncertainty  is represented using {\em possibility} functions, rather than {\em probability} distributions. Possibility functions model the uncertainty in a non-additive manner, and have the capacity to deal with partial (incomplete) problem specification. Thus, the main advantage of the possibilistic Bernoulli filter, derived in this paper, is that it can operate even in the absence of precise measurement and/or dynamic model parameters. This feature of the proposed filter is demonstrated in the context of target tracking using  multi-static Doppler shifts as measurements.
\end{abstract}

\begin{keyword}
Target tracking; Possibility functions; Partially known probabilistic models.
\end{keyword}

\end{frontmatter}

\newpage

\section{Introduction}

Estimation of stochastic dynamic systems (stochastic filtering) is typically carried out using the sequential Bayesian estimation framework  \cite{jazwinski_70}. Assuming the dynamic system is fully characterised by its (hidden) state, the application of the sequential Bayesian estimation method requires the specification of two stochastic models: the dynamic model, which describes the evolution of the (hidden) state, and the observation model, which specifies the relationship between the sensor measurements and the (hidden) state. Practical applications of Bayes filtering are widespread, including target tracking,  communications, navigation, field robotics, bio-informatics, finance, ecology, etc.

In situations where the dynamics and/or observation models are only partially known, using the sequential Bayesian estimation method is not straightforward. If, for example, some parameters of the model(s) are unknown, one approach would be to estimate (learn) their values sequentially from the data \cite{ristic2018measurement}. This, of course, has its limitations, because of limitations in available computational power or observability issues. Different methods such as
Bayesian non-parametric models \cite{ferguson1973bayesian} allow for acknowledging
that all the parameters in the selected dynamical and observation
processes might not be perfectly known, however, these
often involve even more parameters in order to describe what
is the uncertainty on the original ones, thus only offering
a partial solution to the problem.

Research into reasoning under uncertainty in artificial intelligence (AI) is mainly focused on representation and explanation of uncertainty and inference rules for derivation of (uncertain) conclusions. Uncertainty in this context is classified either as aleatory (due to the random effects) or epistemic (due to imprecision, or partial knowledge) \cite{o2004dicing}. The research community in AI has recognised for some time that probability distributions are perfect to represent aleatory uncertainty, but inappropriate to capture effectively the uncertainty caused by ignorance, imprecision or partial knowledge \cite{klir2005uncertainty,denoeux_2017}. Alternative modelling of uncertainty have been proposed by different generalisations
of probability theory, such as fuzzy logic \cite{zadeh1965fuzzy}, imprecise probabilities \cite{walley1991statistical}, possibility theory \cite{zadeh_78,Dubois2015} and Dempster-Shafer theory \cite{Dempster1967,shafer1976mathematical}. Most of these approaches offer
the ability to model a complete absence of information, but do
not provide a general way of dealing with stochastic filtering. In addition, reasoning under uncertainty in AI is typically restricted to discrete state spaces.

In this paper we develop a stochastic filter for joint detection and state estimation of a dynamic object in the presence of false and miss detections, using exclusively {\em possibility} functions. The filter is referred to as the {\em possibilistic Bernoulli filter} (PBF), because it is the possibility theoretic analogue of the standard Bernoulli filter \cite{ristic2013tutorial}, originally derived by Mahler \cite{mahler_07} using probability distributions and random finite sets.  The motivation for using possibility functions, as non-additive models of uncertainty \cite{hampel2009nonadditive}, instead of the probabilistic framework, is to provide an alternative representation of uncertainty, capable of handling, in a rigorous mathematical manner, the situations of ignorance or partial knowledge. Derivation of the PBF follows from the recently proposed framework for stochastic filtering using a class of outer measures  \cite{houss_18,houssineau2017sequential,bishop2018spatio,iet_2019}. Bayes filtering style analytic expressions for prediction and update of outer measures have been formulated and implemented using numerical approximations, such as the grid-based and sequential Monte Carlo (SMC) methods.

The paper is organised as follows. Following \cite{iet_2019}, Sec.~\ref{s:2} reviews the standard Bayes filter and the possibilistic stochastic filter. The PBF is derived in Sec.~\ref{s:3} for the multi-sensor case and an application to target tracking using multi-static Doppler measurements is presented in Sec.~\ref{s:5}. The emphasis in this application is that the probability of detection of each sensor is only partially known, that is, as an interval value. The findings in this article are summarised in Sec.~\ref{s:6}.

\section{Background}
\label{s:2}

\subsection{The standard Bayes filter}

The  stochastic filtering problem in the
Bayesian framework can be formulated as follows \cite{jazwinski_70}. Let us introduce a random variable $\xb_k\in\mathcal{X}\subseteq\mathbb{R}^{n_x}$, referred to as the state-vector, as  the complete
specification of the state of a dynamic system
at time $t_k$. Here $\mathcal{X}$ is the
state space, while $k$ is the discrete-time index corresponding to
time $t_k$. The problem is specified with two equations \cite{pfbook}:
\begin{align}
        \xb_{k} & = \fb_{k-1}(\xb_{k-1})+ \vb_{k-1},   \label{e:dyn_model}\\
        \zb_{k} & = \hb_k(\xb_{k}) + \wb_{k},    \label{e:meas_model}
\end{align}
referred to as the {\em dynamics equation} and the {\em observation
 (or measurement) equation}, respectively. The function $\fb_{k-1}:\mathbb{R}^{n_x}
\rightarrow \mathbb{R}^{n_x}$ is a nonlinear transition function
defining the evolution of the state vector as a first-order Markov
process. The random process  $\vb_{k}\in\mathbb{R}^{n_x}$  is
independent identically distributed~(IID) according to the probability density function (PDF)
$p_{\vb}$; and $\vb_{k}$ is referred to as {\em process noise}. Its
role is to model random disturbances affecting the state evolution model.  The function $\hb_k:\mathbb{R}^{n_x} \rightarrow
\mathbb{R}^{n_z}$ defines the relationship between the state $\xb_k$
and the measurement $\zb_{k}\in\mathcal{Z}$, where
$\mathcal{Z}\subseteq\mathbb{R}^{n_z}$ is the measurement space.
The random process $\wb_k\in\mathbb{R}^{n_z}$, independent of $\vb_k$,
is also IID with PDF $p_{\wb}$, and referred to as {\em measurement
noise}.

In the formulation
(\ref{e:dyn_model})-(\ref{e:meas_model}), functions $\fb_k$ and
$\hb_k$, as well as PDFs $p_{\vb}$ and $p_{\wb}$ are known.
Equations (\ref{e:dyn_model}) and (\ref{e:meas_model}) effectively
define two probability distributions, the {\em transitional density}
$p_{k|k-1}(\xb_k|\xb_{k-1})$  and the likelihood function
$\ell_k(\zb_k|\xb_k)$, respectively. Given the transitional density, the likelihood function, and the initial density of the state (at $k=0$), $p_0(\xb_0)$, the
goal of stochastic Bayesian filtering  is to estimate recursively
the posterior PDF of the state, denoted $p_{k|k}(\xb_k|\zb_{1:k})$, where
$\zb_{1:k} \overset{\mbox{\tiny{abbr}}}{=} \zb_{1},\zb_2,\dots,\zb_k$.

The solution is usually presented as a two step procedure. Let
$p_{k-1|k-1}(\xb|\zb_{1:k-1})$ denote the posterior PDF at $k-1$. The
first step {\em predicts} the density of the state to (the future) time $k$  via the Chapman~-~Kolmogorov
equation \cite{jazwinski_70}:
\begin{equation}
p_{k|k-1}(\xb|\zb_{1:k-1}) = \int
p_{k|k-1}(\xb|\xb')p(\xb'|\zb_{1:k-1}) d\xb'.
\label{e:pred_eq}
\end{equation}
The second step applies the Bayes rule to {\em update} the predicted
PDF using a measurement $\zb_k$ which becomes available at time $k$:
\begin{equation}
p_{k|k}(\xb|\zb_{1:k}) = \frac{\ell_k(\zb_k|\xb)\,
p_{k|k-1}(\xb|\zb_{1:k-1})}{\int \ell_k(\zb_k|\xb')\, p_{k|k-1}(\xb'|\zb_{1:k-1})
d\xb'}. \label{e:upd_eq}
\end{equation}
Knowing the posterior $p_{k|k}(\xb|\zb_{1:k})$, one can compute the point
estimates of the state, such as the expected a posterior (EAP) estimate
or the maximum a posterior (MAP) estimate.

\subsection{The possibilistic stochastic filter}
\label{s:psf}

Instead of random variables, we now consider {\em uncertain variable} \cite{houssineau2018parameter} in order to enable imprecision of the probabilistic model to be considered (epistemic uncertainty). Uncertain variables can be described by outer measures of a certain form, however we consider the case where all the uncertainty is modelled as epistemic uncertainty, so that these outer measures simplify to possibility measures, as introduced in the seminal paper \cite{zadeh_78}.


Let $\Ab$ be a subset of the state space $\mathcal{X}$ and let $\Pi$ be a possibility measure associated with $\xb\in\mathcal{X}$. Then the possibility measure of $\Ab$ takes a value in the interval $[0, 1]$ representing the possibility that $\xb\in\Ab$, and is defined as $\Pi(\Ab) =\sup_{\xb\in \Ab} \pi(\xb)$, where $\pi(\xb)$ is the possibility function (or distribution \cite{dubois2006possibility}) corresponding to $\Pi$.  The possibility function $\pi:\mathcal{X}\rightarrow [0,1]$ is the
primitive object of possibility theory \cite{dubois2006possibility,dubois2015possibility}, which assigns to each $\xb\in\mathcal{X}$  a degree of possibility of being the true value of the state. It is normalised in the sense that $\sup_{\xb\in \mathcal{X}} \pi(\xb) = 1$. The possibility function can be seen as a membership function determining the fuzzy restriction of minimal specificity\footnote{In the sense that any hypothesis not known to be impossible cannot be ruled out.} about  $\xb$ \cite{zadeh_78}.

Any bounded PDF $p(\xb)$ can be transformed into a possibility function $\pi(\xb)$, and conversely, any integrable possibility function can be transformed into a PDF. An example of transformations is:
\begin{eqnarray}
\pi(\xb) & = & \frac{p(\xb)}{ \sup_{\xb'\in \mathcal{X}} p(\xb')},  \label{e:conv1}\\
p(\xb) & = & \frac{\pi(\xb)}{ \int_{\mathcal{X}} \pi(\xb') d \xb'}. \label{e:conv2}
\end{eqnarray}
Other transformations are discussed in \cite{dubois2015possibility}.

Considering that in majority of applications, both process noise and measurement noise are modelled by Gaussian distributions, we can focus on the Gaussian possibility function:
\begin{equation}
\bar{\mathcal{N}}(\xb;\muB,\Pb) = \exp\left(-\frac{1}{2}(\xb-\muB)^\intercal\Pb^{-1}(\xb-\muB)\right)
\label{e:Gauss}
\end{equation}
for some $\muB\in\mathbb{R}^d$ and for some $d\times d$ positive definite matrix $\Pb$ with real coefficients. With abuse of language, we will refer to $\muB$ and $\Pb$ as to the mean\footnote{The possibilistic mean value has been defined as a closed interval \cite{carlsson2009possibilistic}, although other interpretations exist.}
 and the covariance matrix of the Gaussian possibility function $\bar{\mathcal{N}}(\xb;\muB,\Pb)$.

 The goal of the possibilistic stochastic filter is to estimate sequentially the posterior possibility function $\pi_{k|k}(\xb|\zb_{1:k})$. Suppose the posterior possibility function at time $k-1$, $\pi_{k-1|k-1}(\xb|\zb_{1:k-1})$, is available. The prediction equation explains how to compute the possibility function of the state at time $k$ using the transitional possibility function $\rho_{k|k-1}(\xb_k|\xb_{k-1})$, and is given by  \cite{houss_18}:
\begin{equation}
\pi_{k|k-1}(\xb|\zb_{1:k-1}) = \sup\limits_{\xb'\in\mathcal{X}} \rho_{k|k-1}(\xb|\xb')\,\pi_{k-1|k-1}(\xb'|\zb_{1:k-1}).
\label{e:pred_eq_pf}
\end{equation}
The transitional possibility function $\rho_{k|k-1}(\xb|\xb')$ can be, for example, obtained from $p_{k|k-1}(\xb|\xb')$ using transformation (\ref{e:conv1}).
Note that (\ref{e:pred_eq_pf}) is an analogue
of the Chapman-Kolmogorov equation of the standard
Bayes filter (\ref{e:pred_eq}). The two expressions differ only as follows: (i) the integral in (\ref{e:pred_eq}) is replaced by the
supremum in (\ref{e:pred_eq_pf}); (ii) the PDFs in (\ref{e:pred_eq}) are replaced with
possibility functions  in (\ref{e:pred_eq_pf}).

The update step of the possibility filter ``corrects''  the predicted possibility function  $\pi_{k|k-1}(\xb|\zb_{1:k-1})$ using the information contained in the new measurement $\zb_k$. The update equation is given by  \cite{houss_18}:
\begin{equation}
\pi_{k|k}(\xb|\zb_{1:k}) = \frac{g(\zb_k| \xb)\,\pi_{k|k-1}(\xb|\zb_{1:k-1})}{\sup_{\xb\in\mathcal{X}} \left[g(\zb_k|\xb)\,\pi_{k|k-1}(\xb|\zb_{1:k-1})\right]},
\label{e:upd_eq_pf}
\end{equation}
where $g(\zb|\xb)$ represents the likelihood function expressed as a possibility function.
 Note that (\ref{e:upd_eq_pf}) is an analogue of the Bayes' update (\ref{e:upd_eq}). Again, the two expressions differ only in the following: (i) the supremum replaces the integral; (ii) the probability distributions are replaced with the possibility functions.

It was demonstrated in \cite{houss_18} that the predicted and posterior mean and variance in the recursion \eqref{e:pred_eq_pf}-\eqref{e:upd_eq_pf} are the ones of the Kalman filter in the linear-Gaussian case. In the non-linear case, a comparison (in the context of bearings-only tracking) between the standard Bayes filter and the possibilistic stochastic filter \cite{iet_2019} revealed that: (a) in the absence of a model mismatch, the two filters perform identically; (b) in the presence of a (dynamic or observation) model mismatch, the possibilistic filter consistently results in a lower probability of divergence, which indicates a more robust performance.

 \section{Formulation of the PBF}
 \label{s:3}

The Bernoulli filter is a Bayes-type filter, designed for dynamic systems that are capable of switching on and off. In the target tracking context, this means that the target can appear/disappear from the region of interest. The Bernoulli filter estimates recursively the posterior probability of target presence, in addition to the posterior density of the target state.  By monitoring the probability of target presence, one can effectively {\em detect} when the target appears or disappears.

Derivation of the Bernoulli filter\footnote{Which generalises the integrated probabilistic data association (IPDA) filter of \cite{musicki1994integrated}.} \cite{mahler_07,ristic2013tutorial}  was carried out in the framework of random finite set (RFS) theory. By replacing the concept of a random variable with an uncertain variable \cite{houss_18},  and a concept of a RFS with an uncertain finite set (UFS), referred to as an {\em uncertain counting measure} in \cite{houssineau2018detection}, we will next derive a possibilistic analogue of the standard Bernoulli filter.

\subsection{Uncertain finite sets}

An UFS $\Xb\in\mathcal{F(X)}$ is an uncertain variable that takes values as unordered finite sets \cite{mahler_07} on $\mathcal{X}$.  Here $\mathcal{F(X)}$ denotes the set of all finite subsets of $\mathcal{X}$. Both cardinality and the spatial distribution of the elements of $\mathcal{X}$ are uncertain. An UFS is completely characterised by:
 \begin{description}
 \item{(i)} a cardinality distribution, modelled by a discrete possibility function $c(n) = \Pi\{|\Xb|=n\}$, where $n\in\mathbb{N}_0$, $\Pi\{A\}$ is a possibility of event $A$. Due to normalisation, $\max_{n\geq 0} c(n) = 1$;
 \item{(ii)} a family of symmetric possibility functions  $\pi_n(\xb_1,\dots,\xb_n)$, with $n\in\mathbb{N}_0$, $\xb_1,\dots,\xb_n\in\mathcal{X}$. Due to normalisation, $\sup_{\xb_1,\dots,\xb_n} \pi_n(\xb_1,\dots,\xb_n)=1$.
     \end{description}
 The possibility function of an UFS is defined as:
\begin{equation}
f(\{\xb_1,\dots,\xb_n\}) = c(n) \pi_n(\xb_1,\dots,\xb_n)
\label{e:pd_ufs}
\end{equation}
for $n\in\mathbb{N}_0$. For a special class of UFSs, with a property  that points $\xb_1,\dots,\xb_n$ are independently described by a single possibility function $\pi(\xb)$, the possibility function can be expressed as:
\begin{equation}
f(\Xb) = c(|\Xb|) \prod_{\xb \in\Xb} \pi(\xb).
\label{e:pd_ufs1}
\end{equation}

\begin{figure}[htb]
  \centerline{\includegraphics[width=6.cm]{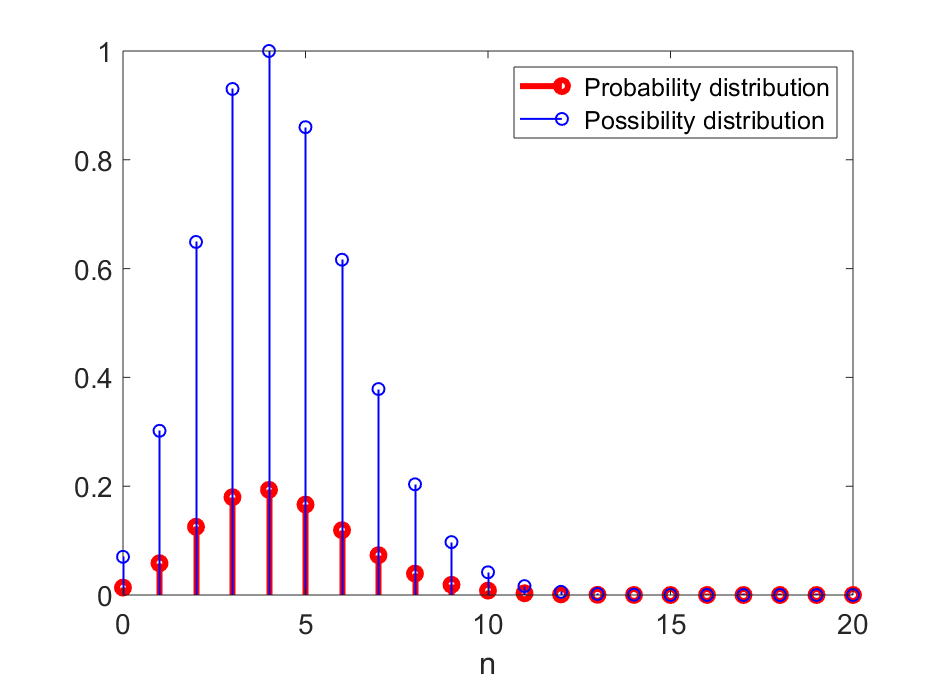}}
  \caption{Illustration of the Poisson probability distribution and possibility function (parameter $\lambda = 4.2$)}
  \label{f:poiss}
\end{figure}

 \paragraph{Example} A Poisson discrete possibility function, with a parameter $\lambda> 0$, can be obtained simply as follows:
 \begin{equation}
 c(n) = \frac{1}{\beta} \frac{\lambda^n \, e^{-\lambda}}{n!}
 \end{equation}
 where
 \begin{equation}
 \beta = \max_{n\geq 0} \frac{\lambda^n \, e^{-\lambda}}{n!} = \frac{\lambda^{ \lfloor \lambda \rfloor} \, e^{-\lfloor \lambda \rfloor}}{\lfloor \lambda \rfloor!}
 \end{equation}
 because $\lfloor \lambda \rfloor$ is the mode of this distribution.  Fig.~\ref{f:poiss} illustrates the Poisson probability (red) and possibility function (blue), for parameter $\lambda=4.2$. $\blacksquare$

As a possibility function on $\mathcal{F}(\mathcal{X})$, $f(\cdot)$ has a supremum equal to one on this domain. This is satisfied by construction since
\begin{align*}
\sup_{\Xb \in \mathcal{F}(\mathcal{X})} f(\Xb) & = \max\Big\{c(0),\,c(1)\sup_{\xb_1 \in \mathcal{X}} \pi_1(\xb_1),\, c(2)\sup_{(\xb_1,\xb_2) \in \mathcal{X} \times \mathcal{X}} \pi_2(\xb_1,\xb_2), \dots\Big\} \\
& = \max_{n \geq 0} c(n) = 1. 
\end{align*}

A Bernoulli UFS $\Xb$ is a set whose cardinality can be 0 or 1. Its UFS possibility function is then:
\begin{equation}
f(\mathbf{X}) = \begin{cases} q^0, & \mbox{if } \mathbf{X} = \emptyset\\
                            q^1\pi(\xb), & \mbox{if }  \mathbf{X} = \{\xb \},
                                \end{cases}
                                \label{e:bern}
\end{equation}
where
\begin{itemize}
\item $q^0=c(0)$ is the possibility that $\mathbf{X}=\emptyset$,
\item $q^1=c(1)$ is the possibility that  $|\mathbf{X}|=1$,
\item $\pi(\xb)$ is the possibility function over $\mathcal{X}$, given that $|\mathbf{X}|=1$.
\end{itemize}
Due to normalisation, it holds that $\max\{q^0,\,q^1\sup_{\xb\in\mathcal{X}}\pi(\xb)\} = 1$. By definition, it also holds that $\sup_{\xb\in\mathcal{X}}\pi(\xb)=1$ and thus the normalisation constraint is simply $\max\{q^0,q^1\}=1$.

\subsection{Dynamic model and the prediction step}
\label{s:a}

If the target is present at time $k-1$ and $k$, then its dynamics is characterised by
the transitional possibility function $\rho_{k|k-1}(\mathbf{x}|\mathbf{x}')$, introduced earlier, see eq. (\ref{e:pred_eq_pf}).  In order to model object appearance and disappearance, it is convenient to introduce a binary uncertain variable $\epsilon_k\in\{0,1\}$ referred to as the target \emph{presence}. The
convention is that $\epsilon_k=1$ means that the target is present at $k$ (and conversely,  $\epsilon_k=0$ means that it is absent). Let the dynamics of $\epsilon_k$ be modelled by a two-state Markov chain with a (time-invariant) transitional possibility matrix (TPM)
\begin{equation}
\mathcal{T} = \left[%
\begin{matrix}
\tau_{\zero\zero} & \tau_{\zero\one} \\
\tau_{\one\zero} & \tau_{\one\one}\end{matrix}%
\right]
\end{equation}
where $\tau_{ij}$ is the possibility of transition from $\epsilon_{k-1}=i$ to $\epsilon_{k}=j$, for $i,j\in\{0,1\}$. Due to normalisation, $\max\{\tau_{i0},\tau_{i1}\} = 1$, for $i=0,1$.
We also need to specify the initial possibility (at $k=0$) of target absence and presence, i.e.\ $q^0_0$ and $q^1_0$, respectively, such that $\max\{q^0_0,q^1_0\}=1$. If a target appears at time $k$, the possibility function describing the possibility of its appearance over the state space $\mathcal{X}$ is denoted by $b_{k|k-1}(\xb)$.

Next we introduce the transitional possibility function of a Bernoulli UFS, from time $k-1$ to $k$. Let us denote this possibility function as $\phi_{k|k-1}(\Xb|\Xb')$. If the target was not present at time $k-1$, then
\begin{equation}
\phi_{k|k-1}(\Xb|\emptyset) = \begin{cases} \tau_{\zero\zero}, & \mbox{if } \Xb_k = \emptyset\\
                                      \tau_{\zero\one} b_{k|k-1}(\xb), & \mbox{if } \Xb=\{\xb\}.
                                      \end{cases}
\end{equation}
If the target was present at time $k-1$ and in state $\xb'\in\mathcal{X}$, then
\begin{equation}
\phi_{k|k-1}(\Xb|\{\xb'\}) = \begin{cases} \tau_{\one\zero}, & \mbox{if } \Xb_k = \emptyset\\
                                     \tau_{\one\one}\rho_{k|k-1}(\xb|\xb'), & \mbox{if } \Xb=\{\xb\}.
                                      \end{cases}
\end{equation}

Let the set of measurements at time $k$ be denoted $\Zb_k$. This set may contain false detections, while the true target detection may be missing due to imperfect target detection process.
The target state at time $k$ is represented by a Bernoulli UFS $\Xb_{k}$. The uncertainty of the target state at $k$ is represented by the posterior possibility function $f_{k|k}(\Xb|\Zb_{1:k})$, where $\Zb_{1:k} \overset{\mbox{\tiny{abbr}}}{=} \Zb_1,\dots,\Zb_k$. In order to simplify notation, we will use abbreviation: $f_{k|k}(\Xb|\Zb_{1:k})\overset{\mbox{\tiny{abbr}}}{=}f_{k|k}(\Xb)$.

The prediction equation of the PBF is then as follows. Suppose the posterior possibility function of a Bernoulli UFS at $k-1$, that is $f_{k-1|k-1}(\Xb)$, is available and expressed according to (\ref{e:bern}) as:
\begin{equation}
f_{k-1|k-1}(\Xb) = \begin{cases} q^0_{k-1|k-1}, & \mbox{if } \mathbf{X} = \emptyset\\
                                 q^1_{k-1|k-1}\,\pi_{k-1|k-1}(\xb), & \mbox{if }  \mathbf{X} = \{\xb \}.
                                \end{cases}
\end{equation}
  Prediction of this possibility function to time $k$ is carried out using the transitional possibility function $\phi_{k|k-1}(\Xb|\Xb')$. Analogue to (\ref{e:pred_eq_pf}), it can be written as:
\begin{equation}
f_{k|k-1}(\Xb) = \sup_{\Xb'\in\mathcal{F(X)}} \left[\phi_{k|k-1}(\Xb|\Xb') f_{k-1|k-1}(\Xb')\right]. \label{e:fffm1}
\end{equation}
When we work out (\ref{e:fffm1}) for $\Xb=\emptyset$, we obtain (see \ref{s:app1}) the prediction equation for the possibility of target being absent at time $k$:
\begin{equation}
q^0_{k|k-1} =  \max\left\{\tau_{\zero\zero}\,q^0_{k-1|k-1},\tau_{\one\zero}\,q^1_{k-1|k-1}\right\}.
\label{e:q0pred}
\end{equation}

By solving (\ref{e:fffm1}) for $\Xb=\{\xb\}$, we obtain (see \ref{s:app1}) the predicted possibility of target being present:
\begin{equation}
q^1_{k|k-1} =  \max \big\{\tau_{\zero\one}\,q^0_{k-1|k-1},
                          \tau_{\one\one}\,q^1_{k-1|k-1}\big\}
\label{e:q1pred}
\end{equation}
and the predicted possibility function over $\mathcal{X}$:
\begin{multline}
\pi_{k|k-1}(\xb) = \frac{1}{q^1_{k|k-1}}  \max \big\{\tau_{\zero\one}\,q^0_{k-1|k-1}\,b_{k|k-1}(\xb), \\
\tau_{\one\one}\,q^1_{k-1|k-1}\,\sup_{\xb'\in\mathcal{X}}[\rho_{k|k-1}(\xb|\xb')\pi_{k-1|k-1}(\xb')]\big\}.
\label{e:piPred}
\end{multline}
It can be easily verified (see \ref{s:app1}) that $\max\{q^0_{k|k-1},q^1_{k|k-1}\}=1$ and that $\sup_{\xb\in\mathcal{X}}\pi_{k|k-1}(\xb)=1$. Also, if the target is present and there are no presence/absence transitions, that is $q^1_{k-1|k-1}=1$, $\tau_{\zero\one}=0$ and $\tau_{\one\one}=1$, then (\ref{e:piPred}) reduces to (\ref{e:pred_eq_pf}).

\subsection{Measurement model and the update step}
\label{s:b}

 Let us assume that $M$ sensors are simultaneously collecting and reporting target measurements. At time $k$, sensor $i\in\{1,\dots,M\}$ reports a (finite) set of measurements (detections) $\Zb^{(i)}_k = \{\zb^{(i)}_{k,1},\zb^{(i)}_{k,2},\dots,\zb^{(i)}_{k,m_k^i}\}$, where both the cardinality of the set $m^i_k\in\mathbb{N}_0$, and the location of the points in $\Zb^{(i)}_k$ in the measurement space $\mathcal{Z}\subset \mathbb{R}^{n_z}$, are uncertain.

 The sensor detector is imperfect in the sense that: (i)  the true target originated measurement may not be present in $\Zb^{(i)}_k$, and (ii) $\Zb^{(i)}_k$ may contain false detections. Suppose the target at time $k$ is in the state $\xb$ and is detected by receiver $i\in\{1,\dots,M\}$, resulting in a measurement $\zb\in\Zb_k^{(i)}$. The likelihood function of $i$th sensor, $g_i(\zb|\xb)$, was introduced in (\ref{e:upd_eq_pf}). It  is expressed as a possibility function over $\mathcal{Z}$ because it specifies the uncertain relationship between the measurement and the target state.

  In accordance with (\ref{e:upd_eq_pf}), the update step of the Bernoulli filter consists of two stages:
   \begin{enumerate}
   \item The predicted Bernoulli possibility function $f_{k|k-1}(\Xb)$ is multiplied with the likelihood function for all measurement sets $\Zb_k \overset{\mbox{\tiny{abbr}}}{=} \Zb^{(1)}_k,\dots,\Zb^{(M)}_k$, given that the target is in the state $\Xb$; this likelihood is denoted $\varphi(\Zb_k|\Xb)$;
   \item Normalisation of the product computed in stage 1.
   \end{enumerate}
Mathematically, the update step can be expressed as:
\begin{equation}
f_{k|k}(\Xb) = \frac{\varphi(\Zb_k|\Xb)\, f_{k|k-1}(\Xb)}{\sup_{\Xb\in\mathcal{F(X)}} \left[ \varphi(\Zb_k|\Xb)\, f_{k|k-1}(\Xb)\right]}
\label{e:bayes_poss_sets}
\end{equation}
where $f_{k|k-1}(\Xb)$, specified by the triplet $\big(q^0_{k|k-1},q^1_{k|k-1}, \pi_{k|k-1}(\xb)\big)$, is in the form (\ref{e:bern}). The terms in the triplet can be computed via (\ref{e:q0pred}), (\ref{e:q1pred}) and (\ref{e:piPred}), respectively.

Next we derive the likelihood function $\varphi(\Zb_k|\Xb)$. Assuming the sensors are independent, we can express this likelihood as a product \cite[Def.4]{shenoy1992using}:
\begin{equation}
 \varphi(\Zb_k|\Xb) = \prod_{i=1}^M \varphi_i(\Zb^{(i)}_k|\Xb).
 \label{e:ind_sens}
 \end{equation}
Note that an UFS $\Zb^{(i)}_k$ (collected by $i$th sensor at time $k$) can be seen as a union of two independent UFSs $\Zb^{(i)}_k = \Cb^{(i)}_k \cup \Wb^{(i)}_k$,  where $\Cb^{(i)}_k$ is the UFS of false detections  and $\Wb^{(i)}_k$ is a Bernoulli UFS modeling the detection from the target \cite{mahler_07}. The target may not be detected, and hence the possibility function of $\Wb^{(i)}_k$, given that target state is $\Xb=\{\xb\}$, according to  (\ref{e:bern}) can we expressed as:
\begin{equation}
\eta_i(\Wb^{(i)}_k|\{\xb\}) = \begin{cases} d_i^0, & \mbox{if } \Wb^{(i)}_k = \emptyset\\
                                 d_i^1 g_i(\zb|\xb), & \mbox{if }  \Wb^{(i)}_k = \{\zb \},
                                \end{cases}
                                \label{e:bern1}
\end{equation}
  where $d_i^0$ and $d_i^1$ denote the possibility of target non-detection and detection by sensor $i$, respectively. Due to normalisation, $\max\{d_i^0,d_i^1\} = 1$.

When the target is absent (i.e.\ $\Xb = \emptyset$), the target originated detection is also absent (i.e.\ $\Wb^{(i)}_k=\emptyset$), hence the possibility function of $\Zb^{(i)}_k$ equals the possibility function of false detections only, given in the form of (\ref{e:pd_ufs1}):
\begin{equation}
\varphi_i(\Zb^{(i)}_k|\emptyset)  = 
\kappa_i(\Zb^{(i)}_k) = \nu_i(|\Zb^{(i)}_k|)\; \prod_{\zb\in\Zb^{(i)}_k}\,\mu_i(\zb). \label{e:var0}
\end{equation}
Here $\nu_i(n)$ is a discrete possibility function of the count of clutter measurements and $\mu_i(\zb)$ is the possibility function on $\mathcal{Z}$ describing the clutter.

If the target is present, the likelihood function $\varphi_i(\Zb^{(i)}_k|\Xb)$  can be expressed as follows:
\begin{subequations}
\begin{align}
\varphi_i(\Zb^{(i)}_k|\{\xb\}) & = \max_{\Wb\subseteq \Zb^{(i)}_k}\left[\eta_i(\Wb|\{\xb\}) \kappa_i(\Zb^{(i)}_k \backslash \Wb)\right] \label{e:varphi1}\\
      & = \max\Big\{\eta_i(\emptyset|\{\xb\}) \kappa_i(\Zb^{(i)}_k), \max_{\zb\in\Zb^{(i)}_k}\big[\eta_i(\{\zb\}|\{\xb\}) \kappa_i(\Zb^{(i)}_k\!\setminus\!\{\zb\})\big]\Big\}  \label{e:varphi2}\\
      & = \max\Big\{d_i^0 \kappa_i(\Zb^{(i)}_k), \max_{\zb\in\Zb^{(i)}_k}\big[d_i^1 g_i(\zb|\xb) \kappa_i(\Zb^{(i)}_k\!\setminus\!\{\zb\})\big]\Big\} \label{e:varphi3}
\end{align}
\end{subequations}
 where the sign $\setminus$ in (\ref{e:varphi2}) and (\ref{e:varphi3}) denotes the set-minus operation. Note that (\ref{e:varphi1}) represents the convolution formula \cite{mahler_07} for UFSs, while (\ref{e:varphi2}) is its simplification because $\eta_i(\Wb|\{\xb\}) = 0$ whenever $|\Wb| > 1$.

 Next we substitute expressions for $\varphi_i(\Zb^{(i)}_k|\emptyset)$ and $\varphi_i(\Zb^{(i)}_k|\{\xb\})$, given by (\ref{e:var0}) and (\ref{e:varphi3}), respectively, in the update equation (\ref{e:bayes_poss_sets}). This leads to the multi-sensor update equations of the PBF (full derivation is given in \ref{s:app4}). The posterior possibility of target absence and presence are given by:
 \begin{eqnarray}
 q^0_{k|k} & = & \frac{q^0_{k|k-1}}{\max\big\{q^0_{k|k-1},\alpha\,q^1_{k|k-1}\big\}}
 \label{e:poss_bern_up_q0}\\
q^1_{k|k} & = & \frac{\alpha \, q^1_{k|k-1}}{\max\big\{q^0_{k|k-1},\alpha\,q^1_{k|k-1}\big\}}
 \label{e:poss_bern_up_q01}
 \end{eqnarray}
respectively, where
 \begin{equation}
 \alpha  = \prod_{i=1}^M R_i(\Zb_k^{(i)})
 \label{e:alpha}
 \end{equation}
 and
 \begin{equation}
 R_i(\Zb)  =
 \max\Big\{d_i^0, d_i^1\max_{\zb\in\Zb}\Big[\frac{\kappa_i(\Zb\!\setminus\! \{\zb\})}{\kappa_i(\Zb)}\,\sup_{\xb\in\mathcal{X}}\big[g_i(\zb|\xb)\pi_{k|k-1}(\xb)\big]\Big]\Big\}.
 \label{e:R}
  \end{equation}
The update equation for the spatial possibility function is:
\begin{equation}
\pi_{k|k}(\xb) = \prod_{i=1}^M \frac{L_i(\Zb_k^{(i)}|\xb)}{R_i(\Zb_k^{(i)})} \; \pi_{k|k-1}(\xb)
\label{e:poss_bern_up_pi}
\end{equation}
where
\begin{equation}
L_i(\Zb|\xb) = \max\Big\{d_i^0,d_i^1\max_{\zb\in\Zb}\Big[\frac{\kappa_i(\Zb\!\setminus\! \{\zb\})}{\kappa_i(\Zb)}\,g_i(\zb|\xb)\Big]\Big\}.
\label{e:LZ}
\end{equation}
It can be easily verified that $\max\{q^0_{k|k},q^1_{k|k}\}=1$ and that $\sup_{\xb\in\mathcal{X}}\pi_{k|k}(\xb) = 1$. Furthermore, consider the case with no false detections, with the possibility of detection  $d_i^1=1$ and the possibility of non-detection  $d_i^0=0$, and with $q^0_{k|k-1}=0$ and $q^1_{k|k-1}=1$. Then $\Zb_k^{(i)}$ contains only one measurement, which must be due to the target.  This leads to $q^1_{k|k} = 1$ and $q^0_{k|k} = 0$, while (\ref{e:poss_bern_up_pi}) reduces to (\ref{e:upd_eq_pf}).

\section{Application: target tracking using multi-static Doppler shifts}
\label{s:5}
\subsection{Problem description}

The problem description follows \cite{ristic2013target}. The state of the moving target in the two-dimensional
surveillance area at time $t_k$ is represented by the state vector
\begin{equation}
\xb_k = \big[\begin{matrix}x_k & \dot{x}_k & y_k &
\dot{y}_k\end{matrix}\big]^\intercal,
\end{equation}
where $^\intercal$ denotes the matrix transpose and $k$ is the discrete-time index. Target position and velocity vector are denoted
$\mathbf{p}_k = [x_k\;\;y_k]^\intercal$ and $\dot{\mathbf{p}}_k = [\dot{x}_k\;\;\dot{y}_k]^\intercal$, respectively. Uncertain target motion is described by the possibility function
\begin{equation}
\rho_{k|k-1}(\xb|\xb') = \bar{\mathcal{N}}(\xb;\Fb\xb',\Qb)
\label{e:trans_poss}
\end{equation}
where
\begin{equation}
\Fb = \Ib_2 \otimes \left[\begin{matrix}1 & T \\ 0 &
1\end{matrix}\right], \hspace{0.3cm} \Qb = \Ib_2 \otimes q
\left[\begin{matrix} \frac{T^3}{3} &
\frac{T^2}{2}\\\frac{T^2}{2} & T\end{matrix}\right].
\end{equation}
Here $\otimes$ is the Kroneker product, $T=t_{k}-t_{k-1}$ is the constant
sampling interval and $q$ is the noise intensity.

\begin{figure}[htb]
\centerline{\includegraphics[height=6cm]{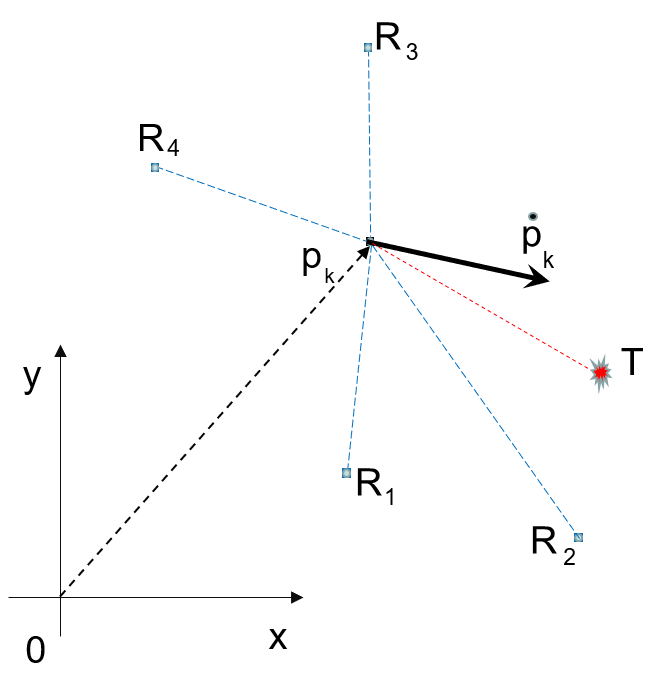}}
 \caption{\footnotesize Multi-static Doppler-only surveillance network: T - transmitter; R$_i$ - $i$th receiver; $\mathbf{p}_k$ and $\dot{\mathbf{p}}_k$  are target position and velocity vector, respectively.}
 \label{f:1}
\end{figure}

 Target tracking is carried out using Doppler shifts measured at spatially distributed receivers, as illustrated  in Fig.~\ref{f:1}. A transmitter T at a known position
$\mathbf{t} = [x_0\;\ y_0]^\intercal$, illuminates the target at
location  $\mathbf{p}_k$ by a sinusoidal waveform of a known carrier
frequency $f_c$. The receivers in Fig.~\ref{f:1} are denoted by
R$_i$, $i\in\{1,\dots,M\}$.

If the target at time $k$ is in the state $\xb_k$, and is detected
by receiver $i\in\{1,\dots,M\}$ placed at a known location
$\mathbf{r}_{i} = [x_i\;\;y_i]^\intercal$, then the receiver will
report a Doppler-shift (measurement) $\zb_k\in \mathcal{Z}=[-f_0,+f_0]$,
described by the likelihood function expressed as a possibility function:
\begin{equation}
g_i(\zb_k|\xb_k) =  \bar{\mathcal{N}}(\zb_k;h_i(\xb_k),\sigma_i^2).
\label{e:lik_msD}
\end{equation}
The frequency $f_0$ (the maximum value of the Doppler shift), is assumed known. The nonlinear measurement function $h_i(\cdot)$ in (\ref{e:lik_msD}) represents the true value of Doppler shift and is given by \cite{ristic2013target}:
\begin{equation}
h_i(\xb_k) = -\dot{\mathbf{p}}_k^\intercal \left[\frac{\mathbf{p}_k
- \mathbf{r}_i}{\parallel\mathbf{p}_k - \mathbf{r}_i
\parallel}
+\frac{\mathbf{p}_k-\mathbf{t}}{\parallel\mathbf{p}_k-\mathbf{t}
\parallel}\right]\frac{f_c}{c}
\label{e:hi}
\end{equation}
where $c$ is the speed of light. In accordance with the comment below (\ref{e:Gauss}), we refer to  $\sigma_i^2$ as to the variance of the Gaussian possibility function (\ref{e:lik_msD}).

The distribution of false detections over the measurement space ${\cal
Z}$ is assumed time invariant and independent of the target state. The number of
false detections per scan is assumed to be Poisson distributed, with
the mean value $\lambda_i$ for receiver~$i$.

Target originated Doppler shift measurement is detected by receiver $i$  with the
probability of detection  $P_{d}^{i}(\xb_k)\leq 1$. In general, the probability of detection is a function of the distance  between the target at position $\mathbf{p}_k$ and the receiver  at location $\mathbf{r}_{i}$, i.e.\
$d^i_k = \| \mathbf{p}_k - \mathbf{r}_i \|$.
For illustration, we adopt a formula  $P^i_d(\xb_k) = \exp[- (d^i_k/\beta)^4]$, where for $d^i_k$ in meters $\beta = 12\cdot 10^3$ (also in meters). Then, $P^i_d$ is a monotonically decreasing function of distance, equaling $1$ at $d^i_{k}=0$, and dropping to $1/2$ at $d_k^i\approx 8320$m. In simulations, the Doppler-shift measurements were generated using this formula for the probability of detection.

We argue that in practice, the probability of detection available to the filter, cannot be as precise as specified above, because in reality it would depend on the signal to noise ratio, which is unknown. For the same reasons, learning the functional form of $P^i_d$, $i=1,\dots,M$, from the data would also be fairly difficult. The main advantage of the PBF over the standard Bernoulli filter \cite{ristic2013tutorial} (formulated using the probability distributions and based on precise specification of all parameters, including $P^i_d$), in this application would be that it needs only a partial knowledge of $P^i_d$, via $d_i^0$ and $d_i^1$, see Sec.~\ref{s:b}. The pair $(d_i^0,d_i^1)$, where $\max\{d_i^0,d_i^1\} = 1$,  effectively defines the interval of detection probability\footnote{The possibility of detection is the upper probability of detection, while the lower probability of detection is the {\em necessity}, defined as one minus the possibility of the complement of detection (i.e. non-detection), see \cite{Dubois2015}.}, that is $P^i_d \in [1-d_i^0,d_i^1]$. In the case of the total ignorance about $P^i_d$, we set $d_i^0=d_i^1 = 1$.

\subsection{Implementation of the PBF}

We developed a computer implementation of the PBF based on an adaptation of the SMC method. Note that one cannot sample directly from a possibility function \cite{houssineau2017sequential}. Instead, for a given possibility function $\pi$, samples must be drawn from a PDF $p$ which is induced by $\pi$. While there is an infinite number of ways one can construct $p$ from $\pi$ (one being (\ref{e:conv2})), the natural solution is the one that results in the least informative $p$. Practical details of an SMC method for possibility functions can be found in \cite{houssineau2017sequential}.

Random samples or particles are propagated over time only as the support points of the posterior possibility function $\pi_{k|k}(\xb)$, mimicking an {\em adaptive grid} over the state space $\mathcal{X}$. The weights, associated with these particles, are computed using the PBF equations (\ref{e:piPred}) and (\ref{e:poss_bern_up_pi}).  Prediction of  the posterior possibilities $q^0_{k|k}$ and $q^1_{k|k}$ is based on the straightforward application of equations (\ref{e:q0pred}) and (\ref{e:q1pred}). In the update step, equations (\ref{e:poss_bern_up_q0}) and (\ref{e:poss_bern_up_q01}), the SMC representation of $\pi_{k|k-1}(\xb)$ is used in the computation of $\alpha$ via (\ref{e:alpha}) and (\ref{e:R}).

The point estimate $\hat{\xb}_{k|k}$ is computed as a weighted mean of the particles approximating $\pi_{k|k}(\xb)$.

\subsection{Numerical results}

The following values were used in simulations. The location of the transmitter: $\mathbf{t} = [0\text{m},\; 0\text{m}]^\intercal$; $M=5$ receivers, placed at $\mathbf{r}_1 = [-8000\text{m},\; 3000\text{m}]^\intercal$, $\mathbf{r}_2 = [-9000\text{m},\; 11000\text{m}]^\intercal$, $\mathbf{r}_3 = [-2000\text{m},\;2000\text{m}]^\intercal$, $\mathbf{r}_4 = [1000\text{m},\; 11000\text{m}]^\intercal$ and $\mathbf{r}_5 = [9000\text{m},\; 9000\text{m}]^\intercal$. Other parameters were: $f_c = 900$ MHz, $T=2$ s, $q=0.1$, $f_0=200$ Hz, $\sigma_i = 2.5$ Hz and $\lambda_i=0.5$ for $i=1,\dots,M$. False detections  were uniformly distributed across $\mathcal{Z}$. The initial target state (at $k=1$): $\xb_1 = [-4000\text{m},\; 30\text{m/s},\; 7000\text{m},\; -12\text{m/s}]^\intercal$. The observation interval is 140 seconds (i.e. $k=1,\dots,70$).

The parameters used in the SMC approximation of the PBF were as follows. The number of particles used was 10000. The target birth distribution $b_{k|k-1}(\xb)=\bar{\mathcal{N}}(\xb;\muB_b,\Pb_b)$, where
the mean is $\muB_b=[0\;0\;0\;0]^\intercal$, that is placed at the location of the transmitter, with zero target velocity. The covariance matrix was set to  $\Pb_b =
\text{diag}[(4\text{km})^2$ $ (30\text{m/s})^2$ $(4\text{km})^2$
$(30\text{m/s})^2]$. Furthermore, $\tau_{\zero\zero}=\tau_{\one\one}=1$ and $\tau_{\zero\one}=\tau_{\one\zero} = 0.01$. The initial possibilities of target presence and absence were set to $q^1_{1|1}=1$ and $q^0_{1|1} = 1$, respectively. This corresponds to the total ignorance about target presence, i.e.\ its probability is in the interval $[0,1]$. A track is confirmed when the difference $q^1_{k|k}-q^0_{k|k} \geq 0.5$, corresponding to the probability of target presence being in the interval $[0.5,1]$.

A single run of the PBF for the described simulation scenario is available, as an avi movie, in the Supplementary material.
Fig.~\ref{f:meas} shows a typical set of Doppler-frequency measurements $Z_{1},Z_2,\dots,Z_{70}$ obtained during a single run. Notice the effect of time-varying probability of detection and false Doppler measurements.
\begin{figure}[htb]
  \centerline{\includegraphics[width=8cm]{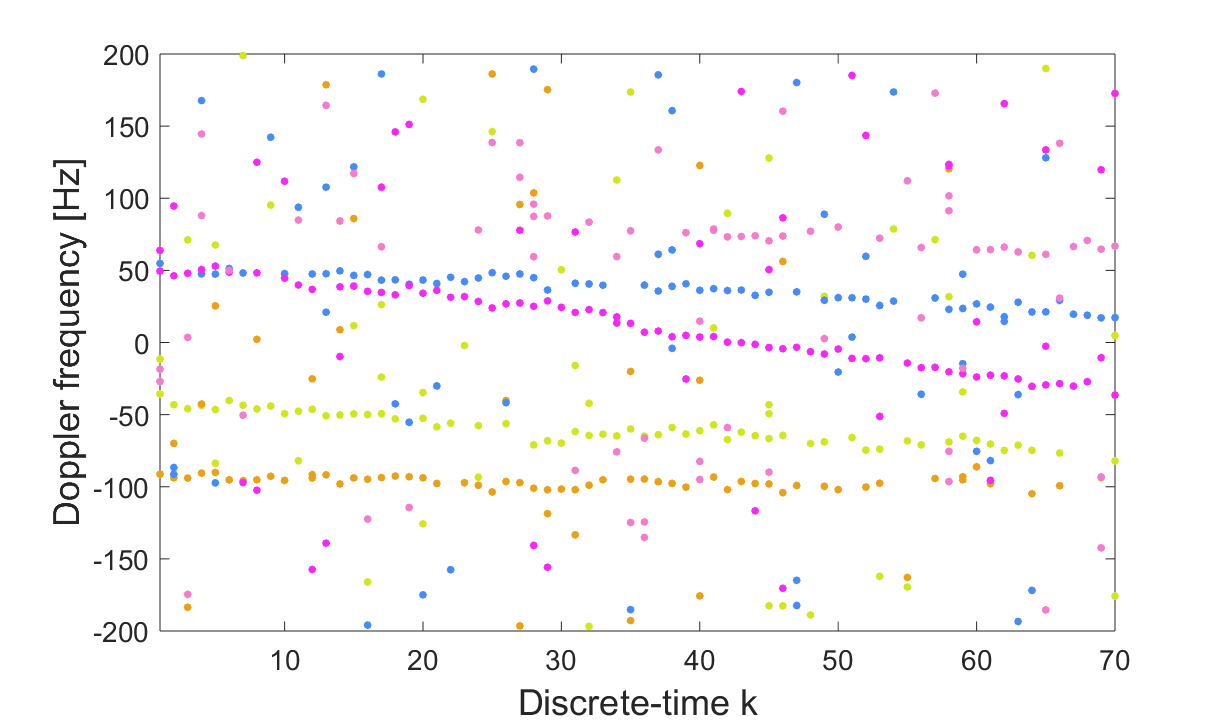}}
  \caption{A typical set of Doppler-shift measurements over time (the same coloured measurements originate from the same receiver)}
  \label{f:meas}
\end{figure}

\begin{figure}[htb]
  \centerline{\includegraphics[width=9.5cm]{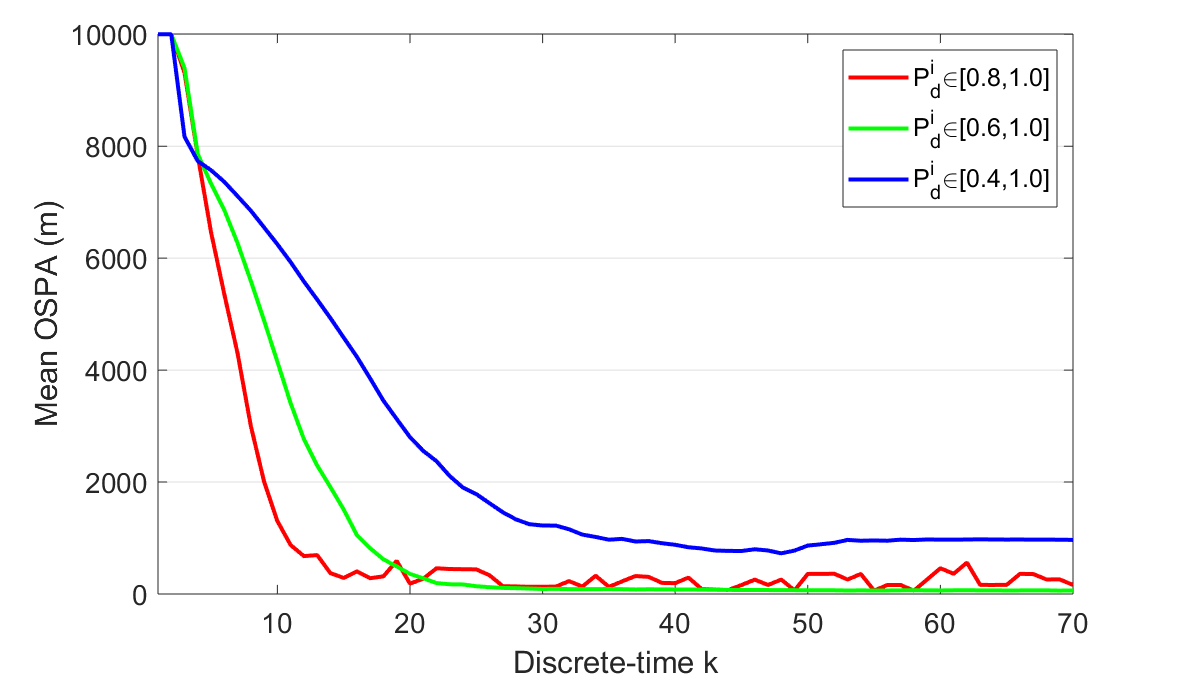}}
  \caption{Mean OSPA error (position) for different specifications of detection probability intervals}
  \label{f:OSPA}
\end{figure}

Fig.~\ref{f:OSPA} presents the mean OSPA errors (in position) \cite{SchuhmacherOSPA07} obtained by averaging over 100 Monte Carlo runs of the PBF. The parameters used in the computation of the OSPA metric were $p=1$ and $c=10^4$m. The three OSPA error curves shown in Fig.~\ref{f:OSPA} correspond to the  three different intervals of probability of detection used in the PBF: blue line for $P_d^i \in [0.4,1.0]$, green line for $P_d^i \in [0.6,1.0]$ and the red line for $P_d^i \in [0.8,1.0]$. Fig. \ref{f:OSPA} demonstrates that the PBF works. The best performance is achieved for  $P_d^i \in [0.6,1.0]$, because this interval captures most accurately the spatio-temporal variation of the probability of detection for all five receivers. By setting $P_d^i \in [0.8,1.0]$, the track is established quicker, however, the track maintenance is less reliable (with occasional breaks in the track). Finally, with $P_d^i \in [0.4,1.0]$, the track is not established in about 10\% of the runs.

\section{Summary}
\label{s:6}

To our best knowledge, the paper presented the first target tracking algorithm completely derived in the framework of possibility theory. The algorithm, referred to as the possibilistic Bernoulli filter, is characterised by Bayesian filtering style analytic expressions for prediction and update. The motivation for using the possibility functions, instead of the probabilistic framework, is a more generalised representation of uncertainty, capable of handling, in a rigorous mathematical manner, the situations of ignorance or partial knowledge. The PBF was demonstrated in the context of an application, where the true (but unknown) probability of detection was varying across the space and time. The PBF was able to track the target using only partial knowledge of the probability of detection, specified as an interval value.

Future research will consider theoretical formulations of other tracking algorithms in the framework of possibility theory.

\appendix
\section{Derivations}

\subsection{Derivation of prediction equations in Sec.~\ref{s:a}}
\label{s:app1}

First we derive equation (\ref{e:q0pred}). Let us start with (\ref{e:fffm1}), i.e.
\begin{align}
f_{k|k-1}(\Xb) & = \sup_{\Xb'\in\mathcal{F}(\mathcal{X})} \left[\phi_{k|k-1}(\Xb|\Xb') f_{k-1|k-1}(\Xb')\right] \nonumber \\
& = \max\Big\{\phi_{k|k-1}(\Xb|\emptyset)  f_{k-1|k-1}(\emptyset),\nonumber \\
& \hspace{3cm} \sup_{\xb'\in\mathcal{X}}\phi_{k|k-1}(\Xb|\{\xb'\}) f_{k-1|k-1}(\{\xb'\})\Big\} \label{e:app1}
\end{align}
For $\Xb=\emptyset$ we have:
\begin{align}
f_{k|k-1}(\emptyset) & =
\max \Big\{\phi_{k|k-1}(\emptyset|\emptyset)  f_{k-1|k-1}(\emptyset), 
\sup_{\xb'\in\mathcal{X}}\phi_{k|k-1}(\emptyset|\{\xb'\}) f_{k-1|k-1}(\{\xb'\})\Big\} \nonumber\\
& = \max\Big\{ \tau_{\zero\zero}\,q^0_{k-1|k-1},\, \tau_{\one\zero}\,q^1_{k-1|k-1} \sup_{\xb'\in\mathcal{X}} \pi_{k-1|k-1}(\xb')\Big\} \label{e:app3}
\end{align}
Note that $\sup_{\xb'\in\mathcal{X}} \pi_{k-1|k-1}(\xb')$, which features on the right-hand side of (\ref{e:app3}), equals to $1$ due to normalisation. Furthermore, since $\Xb$ is a Bernoulli UFS, $f_{k|k-1}(\Xb)$ can be expressed in form (\ref{e:bern}), i.e. as
\begin{equation}
f_{k|k-1}(\Xb) = \begin{cases} q^0_{k|k-1}, & \mbox{if } \mathbf{X} = \emptyset\\
                                 q^1_{k|k-1}\,\pi_{k|k-1}(\xb), & \mbox{if }  \mathbf{X} = \{\xb \}.
                                \end{cases}
                                \label{e:pred_f}
\end{equation}
Hence $f_{k|k-1}(\emptyset)$, which appears on the left hand side of (\ref{e:app3}), represents the predicted possibility that the target is absent, i.e.\ $q^0_{k|k-1}$. Then from (\ref{e:app3}) follows (\ref{e:q0pred}), i.e.
\begin{equation}
   q^0_{k|k-1} = \max\left\{\tau_{\zero\zero}\,q^0_{k-1|k-1},\, \tau_{\one\zero}\, q^1_{k-1|k-1}\right\}.
   \label{e:appq0}
\end{equation}

Next we derive equations (\ref{e:q1pred}) and (\ref{e:piPred}).
First we solve (\ref{e:app1}) for $\Xb = \{\xb\}$:
\begin{align}
f_{k|k-1}(\{\xb\}) & =\max \Big\{ \phi_{k|k-1}(\{\xb\}|\emptyset) f_{k-1|k-1}(\emptyset),\nonumber \\
& \hspace{1.5cm} \sup_{\xb'\in\mathcal{X}}
\big[ \phi_{k|k-1}(\{\xb\}|\{\xb'\}) f_{k-1|k-1}(\{\xb'\}) \big]\Big\} \nonumber \\
&= \max \Big\{\tau_{\zero\one}\,q^0_{k-1|k-1}\,b_{k|k-1}(\xb), \nonumber \\
& \hspace{1.5cm} \tau_{\one\one}\,q^1_{k-1|k-1}\,\sup_{\xb'\in\mathcal{X}}\big[\rho_{k|k-1}(\xb|\xb')\pi_{k-1|k-1}(\xb')\big]\Big\} \label{e:xxx}
\end{align}
From (\ref{e:pred_f}) we have $f_{k|k-1}(\{\xb\}) = q^1_{k|k-1}\pi_{k|k-1}(\xb)$, which leads to:
\begin{equation}
\sup_{\xb \in\mathcal{X}} f_{k|k-1}(\{\xb\}) = q^1_{k|k-1}\,\underbrace{\sup_{\xb \in \mathcal {X}}\pi_{k|k-1}(\xb)}_{=1}
\label{e:trata}
\end{equation}
From (\ref{e:trata}), using (\ref{e:xxx}), we obtain (\ref{e:q1pred}):
\begin{align}
q^1_{k|k-1} &= \sup_{\xb \in \mathcal{X}}\, f_{k|k-1}(\{\xb\}) \nonumber\\
 & =   \max\big\{\tau_{\zero\one} q^0_{k-1|k-1}, \,\tau_{\one\one}q^1_{k-1|k-1}\big\} \label{e:appq1}
\end{align}
using the fact that $\sup_{\xb\in\mathcal{X}}b_{k|k-1}(\xb) = 1$ and $$\sup_{\xb\in\mathcal{\Xb}}\Big[\sup_{\xb'\in\mathcal{\Xb}}\big[\rho_{k|k-1}(\xb|\xb')\pi_{k-1|k-1}(\xb')\big]\Big] = 1.$$
From (\ref{e:pred_f}) we have:
\begin{equation}
\pi_{k|k-1}(\xb)  =  \frac{1}{q^1_{k|k-1}} f_{k|k-1}(\{\xb\}). \label{e:ppp}
\end{equation}
Upon the substitution of $f_{k|k-1}(\{\xb\})$ in (\ref{e:ppp}) with the expression (\ref{e:xxx})  we obtain (\ref{e:piPred}).

Let us show that $\max\{q^0_{k|k-1},q^1_{k|k-1}\}=1$. From (\ref{e:appq0}) and (\ref{e:appq1}) we have:
\begin{align}
&\max\{q^0_{k|k-1},q^1_{k|k-1}\}   \nonumber \\
& = \max\big\{\max\{\tau_{\zero\zero}\,q^0_{k-1|k-1},\, \tau_{\one\zero}\, q^1_{k-1|k-1}\},\max \{\tau_{\zero\one} q^0_{k-1|k-1}, \,\tau_{\one\one}q^1_{k-1|k-1}\} \big\}\nonumber \\
&= \max\big\{\max\{\tau_{\zero\zero}\,q^0_{k-1|k-1},\,\tau_{\zero\one} q^0_{k-1|k-1} \},\max \{\tau_{\one\zero}\, q^1_{k-1|k-1}, \,\tau_{\one\one}q^1_{k-1|k-1}\}\big \}\nonumber \\
& = \max\big\{\underbrace{\max\{\tau_{\zero\zero},\tau_{\zero\one}\}}_{=1} q^0_{k-1|k-1} ,\underbrace{\max\{\tau_{\one\zero},\tau_{\one\one}\}}_{=1}q^1_{k-1|k-1}\big \}\nonumber \\
& = \max\big\{ q^0_{k-1|k-1} ,q^1_{k-1|k-1}\big\}\nonumber \\
& = 1. \nonumber
\end{align}

\subsection{Derivation of update equations in Sec.~\ref{s:b}}
\label{s:app4}

The BPF update equation (\ref{e:bayes_poss_sets}) for independent sensors, see (\ref{e:ind_sens}), is given by:
\begin{equation}
    f_{k|k}(\Xb) = \frac{\prod\limits_{i=1}^M\varphi_i(\Zb_k^{(i)}|\Xb)\,f_{k|k-1}(\Xb)}{\sup\limits_{\Xb\in\mathcal{F(X)}} \left[ \prod\limits_{i=1}^M\varphi_i(\Zb_k^{(i)}|\Xb)\,f_{k|k-1}(\Xb)\right]}
    \label{e:app2_upd}
\end{equation}
where
\begin{equation}
    \varphi_i(\Zb_k^{(i)}|\Xb) = \begin{dcases}\kappa_i(\Zb_k^{(i)}), & \mbox{if } \Xb=\emptyset\\
    \max\Big\{d^0\kappa_i(\Zb_k^{(i)}),
    \max_{\zb\in\Zb_k^{(i)}}\big[d^1\, g_i(\zb|\xb)\kappa_i(\Zb_k^{(i)}\!\setminus\!\{\zb\})\big]\Big\}
    & \mbox{if } \Xb = \{\xb\}\end{dcases}
\end{equation}
and $\kappa_i(\Zb)$ is defined by (\ref{e:var0}). Let us first focus on the denominator of (\ref{e:app2_upd}), which we denote by $D$ for brevity.
\begin{align}
&D = \sup_{\Xb\in\mathcal{F(X)}}\left[ \prod_{i=1}^M\varphi_i(\Zb_k^{(i)}|\Xb)f_{k|k-1}(\Xb)  \right] \nonumber \\
& = \max\left\{\prod_{i=1}^M\varphi_i(\Zb_k^{(i)}|\emptyset)f_{k|k-1}(\emptyset),\;\sup_{\xb\in\mathcal{X}} \prod\limits_{i=1}^M\varphi_i(\Zb_k^{(i)}|\{\xb\})f_{k|k-1}(\{\xb\}) \right\}\nonumber \\
& = \max\Bigg\{q^0_{k|k-1}\prod_{i=1}^M\kappa_i(\Zb_k^{(i)}),\; \sup_{\xb\in\mathcal{X}}\bigg[ q^1_{k|k-1}\pi_{k|k-1}(\xb) \nonumber \\
& \hspace{.7cm} \times \prod_{i=1}^M\max\Big\{d^0\kappa_i(\Zb_k^{(i)}),\max_{\zb\in\Zb_k^{(i)}}\big[d^1\, g_i(\zb|\xb)\kappa_i(\Zb_k^{(i)}\!\setminus\! \{\zb\}) \big] \Big\}\bigg]\Bigg\}\nonumber \\
&=\max\Bigg\{q^0_{k|k-1}\prod_{i=1}^M\kappa_i(\Zb_k^{(i)}), \nonumber \\
& \hspace{.7cm}q^1_{k|k-1}\prod_{i=1}^M\max\Big\{d^0\kappa_i(\Zb_k^{(i)}),\max_{\zb\in\Zb_k^{(i)}}\Big[d^1\, \kappa_i(\Zb_k^{(i)}\!\setminus\! \{\zb\})\,\sup_{\xb\in\mathcal{X}}[g_i(\zb|\xb)\pi_{k|k-1}(\xb)]\Big]\Big\}\Bigg\}\nonumber \\
& = \prod_{i=1}^M\kappa_i(\Zb_k^{(i)}) \max\big\{q^0_{k|k-1},\; \alpha\,q^1_{k|k-1}\big\}
\label{e:DDD}
\end{align}
where
\begin{equation}
\alpha = \prod_{i=1}^M\max\left\{d^0,\; d^1\max_{\zb\in\Zb_k^{(i)}}\Big[ \frac{\kappa_i(\Zb_k^{(i)}\!\setminus\! \{\zb\})}{\kappa_i(\Zb_k^{(i)})}\,\sup_{\xb\in\mathcal{X}}\big[g_i(\zb|\xb)\pi_{k|k-1}(\xb)\big]\Big]\right\}
\end{equation}
Let us now write (\ref{e:app2_upd}) for $\Xb=\emptyset$, recalling that $f_{k|k}(\emptyset) = q^0_{k|k}$ and $f_{k|k-1}(\emptyset)=q^0_{k|k-1}$ and using (\ref{e:DDD}):
\begin{equation}
    q^0_{k|k} = \frac{\prod\limits_{i=1}^M \kappa_i(\Zb_k^{(i)}) q^0_{k|k-1}}{\prod\limits_{i=1}^M \kappa_i(\Zb_k^{(i)}) \max\big\{q^0_{k|k-1},\alpha\, q^1_{k|k-1}\big\}} \label{e:q0q}
    \end{equation}
    After canceling the term $\prod_{i=1}^M \kappa_i(\Zb_k^{(i)})$ in (\ref{e:q0q}) we obtain (\ref{e:poss_bern_up_q0}), i.e.
\begin{equation}
    q^0_{k|k} = \frac{q^0_{k|k-1}}{\max\big\{q^0_{k|k-1},\alpha q^1_{k|k-1}\big\}}
\end{equation}

Next we derive (\ref{e:poss_bern_up_q01}). Because $\Xb$, after the update, remains a Bernoulli UFS, $f_{k|k}(\Xb)$ of (\ref{e:app2_upd}) will be in the form given by  (\ref{e:bern}). Thus,
for the case $\Xb=\{\xb\}$, we have $f_{k|k}(\{\xb\}) = q^1_{k|k}\,\pi_{k|k}(\xb)$. In accordance with the reasoning in (\ref{e:trata}):
\begin{align}
    q^1_{k|k} & = \sup_{\xb \in\mathcal{X}} \, f_{k|k}(\{\xb\})\\
        & = \frac{\sup_{\xb \in\mathcal{X}} \left[\prod_{i=1}^M \varphi(\Zb_k^{(i)}|\{\xb\})\right]\,f_{k|k-1}(\{\xb\})}{D}
\end{align}
where $D$ is given by (\ref{e:DDD}). Following the steps in derivation of $D$, it can be easily shown that
\begin{align}
    q^1_{k|k} & = \frac{\prod\limits_{i=1}^M \kappa_i(\Zb_k^{(i)})\;\, \alpha\, q^1_{k|k-1} }{\prod\limits_{i=1}^M\kappa_i(\Zb_k^{(i)}) \;\max\big[q^0_{k|k-1},\; \alpha\,q^1_{k|k-1}\big]}
    \label{e:q1q}
\end{align}
After canceling the term $\prod_{i=1}^M \kappa_i(\Zb_k^{(i)})$ in (\ref{e:q1q}) we obtain (\ref{e:poss_bern_up_q01}), i.e.
\begin{equation}
    q^1_{k|k} = \frac{\alpha\,q^1_{k|k-1}}{\max\big[q^0_{k|k-1},\; \alpha\,q^1_{k|k-1}\big]}.
    \label{e:q1q1q1}
\end{equation}

Finally, (\ref{e:poss_bern_up_pi}) can be obtained from (\ref{e:app2_upd}) in the case $\Xb=\{\xb\}$. Recall that $f_{k|k}(\{\xb\}) = q^1_{k|k}\,\pi_{k|k}(\xb)$ and $f_{k|k-1}(\{\xb\}) = q^1_{k|k-1}\,\pi_{k|k-1}(\xb)$. Then using (\ref{e:DDD}),  (\ref{e:app2_upd}) can be written as:
\begin{equation}
    q^1_{k|k}\,\pi_{k|k}(\xb) = \frac{\prod\limits_{i=1}^M\kappa_i(\Zb_k^{(i)})\; \prod\limits_{i=1}^M L_i(\Zb_k^{(i)}|\xb)  q^1_{k|k-1}\pi_{k|k-1}(\xb)}{\prod\limits_{i=1}^M\kappa_i(\Zb_k^{(i)}) \max\big\{q^0_{k|k-1}, \alpha\,q^1_{k|k-1}\big\}}
\end{equation}
where $L_i(\Zb_k^{(i)}|\xb)$ was defined in (\ref{e:LZ}). After canceling the term $\prod_{i=1}^M \kappa_i(\Zb_k^{(i)})$ and rearranging we have:
\begin{equation}
    \pi_{k|k}(\xb) = \frac{q^1_{k|k-1}}{q^1_{k|k}\max\big\{q^0_{k|k-1}, \alpha\,q^1_{k|k-1}\big\}}\;\prod\limits_{i=1}^M L_i(\Zb_k^{(i)}|\xb)\pi_{k|k-1}(\xb)
    \label{e:piU}
\end{equation}
Using (\ref{e:q1q1q1}) we simplify (\ref{e:piU}) to:
\begin{equation}
     \pi_{k|k}(\xb) = \frac{1}{\alpha}\prod_{i=1}^M L_i(\Zb_k^{(i)}|\xb)\pi_{k|k-1}(\xb)
\end{equation}
which is identical to (\ref{e:poss_bern_up_pi}).

\end{document}